\documentstyle[aps,prbbib,twocolumn]{revtex}

\begin{document}

\title{Robust Insulator-Superconductor transition on solid inert gas and other substrates}
\author{K. Das Gupta, G. Sambandamurthy, Swati S. Soman and N. Chandrasekhar}
\address{Department of Physics, Indian Institute of Science, Bangalore 560 012,
India. }
\maketitle

\begin{abstract}
We present observations of the insulator-superconductor transition in ultrathin films of Bi on amorphous quartz, quartz coated with Ge, and for the first time, solid xenon condensed on quartz. The relative permeability $\epsilon_r$ ranges from 1.5 for Xe to 15 for Ge. Though we find screening effects as expected, the I-S transition is robust, and unmodified by the substrate.  The resistance separatrix is found to be close to $h/4e^2$ and the crossover thickness close to $\rm 25~\AA$ for all substrates. $I-V$ studies and Aslamazov-Larkin analyses indicate superconductivity is inhomogeneous.  The transition can be understood in terms of a percolation model.
\end{abstract}

\vskip0.5cm

\noindent

The insulator-superconductor (I-S) transition has been extensively investigated 
over the last decade, in a variety of systems such as thin films~\cite{havi,beas}
single Josephson junctions~\cite{pent}, arrays~\cite{van}, one dimensional wires~\cite{tink}. 
Values of limiting resistance close to the quantum resistance for pairs in one dimensional wires~\cite{tink}, and two dimensional, nominally homogeneous ultra-thin films~\cite{havi} are reported.  Differing values of the limiting resistance at the transition have been observed~\cite{beas} in different systems, and attributed to structure, i.e. homogeneous and granular films are expected to behave differently.  

A phase-only picture, first proposed by Ramakrishnan~\cite{tvr}, and further elaborated by Fisher~\cite{fish} has been considered appropriate for such systems. A scaling theory of the I-S transition has been developed.~\cite{mccha}  This theory predicts that the critical resistance will be universal, i.e. independent of all microscopic details, if the system is invariant under the interchange of the roles of charge and flux.  Thus the I-S transition may be self-dual~\cite{wen}.  The precise value of the critical resistance appears to depend on the nature of the interaction between charges, being equal to $h/4e^2$ only when the interaction is logarithmic in the separation, one of the conditions which must exist for self-duality.  The interaction between charges can be logarithmic in their separation if the 2D films have a sufficiently high dielectric constant.~\cite{kel} This condition may be met for semiconductor and semimetal films.     

This has to be reconciled with experimental observations, which show that superconductivity in ultrathin films can be enhanced by the proximity of a metal or dielectric.~\cite{adk}  This is attributed to partial screening of the Coulomb interaction between conduction electrons in the films.  Screening also affects the properties of insulating films, by changing the localization length.~\cite{adk1,ewo}  Glover studied the effect of the substrate dielectric constant on transition temperature of thin films, over thirty years ago, inconclusively.  Ge or Sb has been extensively used as underlayers in ultrathin film studies~\cite{havi,stro,dyne}.  A natural question arises at this point about the effect of the substrate/underlayer on the I-S transition.

For a film on a underlayer, the general expression for the interaction potential between charges in the film is
\begin{equation}
U(r) = {e^2\over{4\pi\epsilon_0\epsilon_r}}{\left({{1\over{r}} - {1\over{\sqrt{r^2+4d^2}}}}\right)}
\end{equation}
where $d$ is the screening distance (the distance between the midplane of the film and the midplane of the underlayer) and $r$ the separation between the charges in the film.  If the separation of two charges is small compared with the distance between a charge and its image ($2d$), the underlayer makes little difference, and the interaction remains monopolar.  At large distances, however the charge and its image behave as a dipole and interaction falls off more rapidly with distance.  For the films studied here, $2d$ would be close to the film thickness.  Although screening by charge carriers in the film is considered in the usual BCS treatment, the relative permeability for a highly disordered film is unknown.~\cite{adams}  It is expected to lie between the metallic and insulating limits of $\epsilon_r\simeq\infty$ and $\epsilon_r\simeq10$.  This issue becomes important in the I-S transition region. $U(q)$, the term in the BCS interaction Hamiltonian becomes
\begin{equation}
U(q) = {4\pi e^2\over{{q^2+k_s^2}}}
\end{equation}
where $k_s$ is the inverse screening length. In principle a film with a transition temperature of 0 K defines $R_c$.  As we show, screening influences the transition temperature of superconducting films, and the normal state resistance of all films. Therefore it should affect the I-S transition.   

It has also been proposed that the I-S transition in the limit $T=0$ is a combined effect
of pairing and localization in 2D systems.~\cite{pang}  Experiments have refuted this conjecture.\~cite{kagawa}  How disorder, cluster size, and the Coulomb interaction influence the behavior of Cooper pairs is still not very clear.  It has been conjectured that the Coulomb interaction may not influence the critical normal state sheet resistance.~\cite{pang}  However, this is refuted by observations of the M-I transition in silicon MOSFETS and other systems in zero magnetic field.~\cite{elihu}  It has been shown that interactions may increase or decrease the conductance of a disordered 2D electron system - weak interactions increase the dc conductance in the localized regime while they decrease the conductance in the diffusive regime.~\cite{vojta} Strong interactions were always found to decrease the conductance.  These considerations motivated our experiments.  

In superconductors, Coulomb interactions suppress the fluctuations in the number of electrons, and increase the fluctuations of the phase of the superconducting order parameter.  This affects the Josephson coupling energy $E_J$ in inhomogeneous systems, where the competition between the charging energy $E_C$ and $E_J$ drives the transition.\~cite{kagawa}  Typical systems exhibiting such behavior are granular films, and junction arrays. $E_J$ may be estimated using the Ambegaokar-Baratoff equation, $E_J=\pi\hbar\Delta/4e^2 R_N$, where $\Delta$ is the BCS gap for Bi, $R_N$ is the normal state resistance of the film. $E_C$ can be estimated as $E_C=q^2/4\pi\epsilon_0\kappa d$ where $\kappa = \epsilon_r(d/2s + 1)$, where $\epsilon_r$ is the relative permeability of the substrate, $d$ is the size of the grains/clusters, and $s$ is the spacing.  Since $R_N$, and $\kappa$ depend on the substrate/underlayer, one expects the I-S transition to occur at different $R_N$ on different substrates.  Since the effects of disorder, cluster size, and interactions on Cooper pairs are unclear, exlpicit predictions of $R_N$ for the I-S transition on various substrates cannot be made.        

In this work, we report studies of the I-S transition in quench condensed Bismuth films, as a function of disorder (or film thickness) on a variety of substrates - amorphous quartz, quartz coated with Ge, and solid xenon condensed on quartz. The relative permeability $\epsilon_r$ ranges from 1.5 for Xe to 15 for Ge. Despite screening effects the I-S transition is robust.  Studies on other substrates such as single crystal sapphire, and sapphire coated with Ge, Xe yield similar results. The experiments were done in a custom UHV cryostat equipped with reflection electron diffraction (RHEED), capable of a hydrocarbon free vacuum of $\sim$ 5$\times$10$^{-10}$ Torr, described elsewhere~\cite{sam1}. The Ge underlayers were deposited on the substrates in a separate UHV system at room temperature.  The Xe underlayers were grown in-situ.  Our experimental setup resembles that of Ref.[1-2], although we cannot attain such low temperatures. We quench condense our ultra-thin films in the temperature range 1.8 K to 15 K. The substrate temperature influences the disorder in the film, and thereby its properties.  Here we report results on films that were quench condensed at 15 K, which facilitates comparison with published results~\cite{havi}.

RHEED studies show that the Bi is almost amorphous.  It is difficult to unambiguously distinguish between amorphous and nanocrystalline at such low temperatures, on poorly conducting films due to charging effects.  Based on the Scherer formula for the peak broadening, we estimate that films thicker than 10 $\rm \AA$ are composed of clusters that vary in size from 25 $\rm \AA$ to 100 $\rm \AA$~\cite{sam1}. Since the information obtained is in reciprocal space, it is difficult to comment on the real space surface morphology.  Our RHEED observations are consistent with previous results.~\cite{vall}  Superconductivity in granular systems of rhombohedral Bi clusters has also been reported.~\cite{mick}

Figure 1 shows the evolution of the temperature dependence of the sheet resistance
R(T) with thickness for (a) Bi films on Ge (10 $\AA$ thick), which has been deposited on amorphous quartz, (b) Bi films on quartz, and (c) Bi films on solid xenon condensed on amorphous quartz.  A transition from insulating type behavior, to superconducting behavior as the thickness of the films is increased is clear.  This type of zero field  transition is considered a zero temperature quantum phase transition, controlled either by disorder, carrier concentration, or thickness.  The normal state resistance at an arbitrarily high temperature $\rm R_N$ has traditionally been used to parameterize the transition, although it may be weakly temperature dependent above the superconducting transition temperature, and becomes ill defined as the I-S transition is approached.  The value of the normal state resistance of a film on the boundary between superconducting and insulating behavior has been referred to as the resistance separatrix, and has been denoted by $\rm R_0$~\cite{havi,beas,tink}.  We obtain $\rm R_0$ as an algebraic average of the sheet resistances of the last insulating and the first superconducting films, measured at a relatively high temperature (10 K).  $\rm R_0$ is close to $h/4e^2$ for all three sets of data.  This observation indicates that the value of $\rm R_0$ is substrate independent, and possibly experiment independent.     

The transition temperature for a film of a given thickness is higher on substrates of higher relative permeability, and the normal state resistance is also higher. A 65 $\rm \AA$ film on Xe, quartz and Ge has $T_c$'s of 3.8, 4.2 and 4.42 K respectively.  These results are consistent with published data.~\cite{adk}  This shows that the conductance of a disordered 
film depends on interplay between interaction and disorder.~\cite{vojta}  We caution that since our experiments are limited to 1.8 K, we cannot rule out the possibility that a film which appears to be the last film on the insulating side, may, at lower temperature turn out to be superconducting.~\cite{dyne}.  A plot of conductance vs thickness at different temperatures gives the crossover thickness, which is shown in Fig. 2 for films on Xe.  We find the crossover thickness to be between $\rm 25~to~28 \AA$ for all substrates.

The insulating films follow a behavior that is consistent with the results reported.~\cite{gold}  We find that the conductivities of all our insulating Bi films, on all substrates to be:
\begin{equation}
\sigma(T) = \sigma_0 exp[-(A/T)^x]
\end{equation}
where x changes with film thickness.  For the thinnest films, x is close to 0.75, due to collective hopping~\cite{gold}.  As the thickness is further increased, x reduces to
0.5, the Efros-Shklovskii form~\cite{es}, which describes hopping modified by Coulomb interactions, and finally for the thickest insulating films x is close to 0.33, the Mott value~\cite{mott}.  

The form of the R(T) for these films may lead us to the conclusion that these films
are nominally homogeneous. Such a conclusion is incorrect as we show below. 
Although STM studies of surface morphology~\cite{vall} report that even at 75\% coverage the films are not conducting, hence no evidence for percolative behavior, we find that superconductivity in our films is indeed percolative in nature.~\cite{stro1}  
Aslamazov and Larkin~\cite{al} considered the possibility of fluctuations causing superconductivity.  The total conductivity is given by $\sigma$ = $\sigma_N$ + $\sigma^{'}_{2D}$, where $\sigma_N$ is the normal state dc conductivity, and $\sigma^{'}$ the paraconductivity. Its temperature dependance is similar to that of the magnetic susceptibility at T$>$T$_c$.  They derived the result,
\begin{equation}
\frac{\sigma^{'}_{2D}}{\sigma_n} = \frac{e^2}{16\hbar}\frac{R^{N}_{\Box}}{\tau} = \frac{\tau_o}{\tau}
\end{equation}
where R$^{N}_{\Box}$ is the normal state sheet resistance, $\tau_o$ = 1.52 $\times$ 10$^{-5}$ R$^{N}_{\Box}$ and  $\tau = \frac{T-T_{c}}{T_{c}}$ is the width factor. T$_c$ is the mean field transition temperature.  $\tau$/R$^{N}_{\Box}$ is a constant ($g_{AL}$=e$^2$/16$\hbar$) for all materials. We have evaluated  $\tau$/R$^{N}_{\Box}$ ($g_{exp}$) for various films. A systematic dependence of $g_{exp}$ on the thickness $d$ is shown in Fig. 3. This parameter deviates from $g_{AL}$ for thinner films.  It approaches the AL value ($g_{AL}$) as the thickness is increased.  It is assumed that theory predicts $g_{exp}$ = $g_{AL}$ for all films, independent of microstructure. Both the normal state conductance and paraconductance depend on sample shape. Glover~\cite{glo} has shown that as the microstructure deviates from a ``uniform rectangular slab'', $g_{exp}$ exceeds the AL value.  The thinner the film, higher the disorder, larger are the deviations from a slab geometry, and larger the deviation of $g_{exp}$ from the AL value.  Hence, films close to the transition are inhomogeneous.

Further evidence is in the form of hysteretic I-V curves, shown in Fig. 4.  These curves can be understood in terms of a resistively and capacitively shunted random Josephson junction array model. These I-V's have been discussed in detail in a separate publication~\cite{sam1}. The physical picture is of superconducting islands connected by thin regions of normal metal, which act as the weak links between the islands, thus forming a random JJ array. The distribution of grain/cluster, each with a finite number of electrons, requires a finite N BCS treatment.  This results in a spread in $T_c$~\cite{tsai}, so that at some temperature all regions of the film are not superconducting.  Normal regions exist, and act as weak links.  Areal maps of quasiparticle conductance or superconducting order parameter amplitude can resolve such regions.  

Feigel'man and Larkin~\cite{larkf} and Spivak {\it et al.}~\cite{spiv} have discussed the quantum superconductor-metal transition in a 2D proximity coupled array.  Thickness larger than the coherence length was considered by the former, with the opposite extreme analyzed by the latter. They predict intervention of a normal phase. Limitations on the accessible temperatures prevent our study confirming this. From the $H_{c2}(T)$ curve, we have determined the coherence lengths ($\xi_c$) to be close to the film thickness. This regime merits further theoretical study.  We have evaluated the localization lengths ($\xi_l$), and we find that at the I-S transition, $\xi_l >> \xi_c$, consistent with the results of Ref. [17].    

The effect of strong disorder on the superconducting order parameter amplitude has been investigated within the Bogoliubov-de Gennes framework~\cite{nand}.  The local pairing amplitude  develops a broad distribution with significant weight near zero, as disorder increases.  The density of states showed a finite spectral gap.  Persistence of the gap was found to arise due to the breakup of the system into superconducting islands.  Superfluid density and off-diagonal correlations showed a substantial reduction at high disorder.  Incorporation of phase fluctuations lead to a non-superconducting state.  Our data are consistent with such a picture.  

Our results suggest that a percolation type model, proposed by Meir~\cite{meir} to
explain the M-I transition in 2D systems is most appropriate and merits investigation. Meir suggested the following physical picture for the M-I transition in 2D - potential fluctuations due to disorder define density puddles of size $L_{\phi}$ or larger, within which the electron wave function totally dephases.  Locally between these puddles, transport is via quantum tunneling. Support for this idea comes from fits of the conductance at a high temperature (10 K) for the insulating films which follows a power law, with an exponent of 1.33 characteristic of 2D percolation, as shown in Fig. 5.  As an interesting observation, we also show that the superconducting films also fit this same power law, albeit with a different critical thickness. The critical thickness for the insulating films is the thickness for onset of electrical conduction.  We have no explanation as to why the critical thickness for normal state conduction depends on whether or not the film superconducts at lower temperature, but present this as an interesting observation.  Keeping in mind the model discussed above, it is logical to associate the puddles with superconducting regions and the quantum point contacts with the normal regions.  The competition between the $E_C$ and $E_J$ probably drives the transition.  However, it is intriguing that the transition is robust at $h/4e^2$ on all substrates, despite $E_C$ and $E_J$ being dependent on $\epsilon_r$.  Percolation issues have been studied earlier.~\cite{kapi,tvr1,yaz}

The conditions for 2D coulomb interactions may have been met in this and other experiments.~\cite{havi}  Keldysh~\cite{kel} has shown that this requires the dielectric constant of the film to be much greater than that of the substrate.  In this study, substrate dielectric constants range from 1.5 for solid Xe to 15 for Ge underlayers.  Hall measurements were done for the first superconducting film near the transition, and this yielded a high carrier concentration of the order of $10^{23} cm^{-3}$, which is consistent with that reported in the literature~\cite{buck}, suggesting a high dielectric constant for the films.  We have not done Hall measurements for films on the insulating side, since interpretation of data in the hopping regime is complicated.  Vortices exist in these films, since they are random Josephson junction arrays,~\cite{xia} due to the discrete nature of the array, which requires that the current take convoluted paths, consistent with a transition governed by percolation.  Duality in JJ arrays has been observed before.~\cite{van}  

In conclusion, we have observed robust I-S transition in ultrathin Bi on several substrates. We have presented, for the first time, the I-S transition on solid inert gas underlayers. Although the substrate influences the $T_c$ as well as $R_N$ of the films, the transition itself is independent of the interaction potential between charges. An AL analysis indicates inhomogenous films, contrary to the R(T).  I-V's indicate a percolation type transition.  A model similar to that proposed by Meir for the M-I transition in 2D merits further investigation. Further studies at lower temperatures are needed.    

This work was supported by DST, Government of India.  We thank Drs. Allen Goldman, C. J. Adkins for discussions. KDG thanks CSIR, New Delhi for the financial support through a Junior Research Fellowship.

\newpage
\centerline{\bf FIGURE CAPTIONS}
\noindent{\bf Figure 1} Evolution of R$_\Box$ vs T of Bi on (a) 10 $\rm \AA$ Ge on amorphous quartz
(b) bare quartz and (c) solid Xe condensed on amorphous quartz. The substrate temperature during deposition in all cases was maintained at 15 K. 

\noindent{\bf Figure 2} Plots of resistance vs. thickness at different temperatures showing the crossover thickness of about $\rm 26 \AA$ for Bi films on Ge.  The crossover occurs within $\rm \pm 3\AA$ of this value for all substrates.

\noindent{\bf Figure 3} The parameter $g_{exp}$ which characterizes the amplitude of the fluctuations as a function of film thickness.  It asymptotically approaches the Aslamazov-Larkin value. 

\noindent{\bf Figure 4} I-V of the superconducting films on quartz.  The hysteresis loops show the dissipation due to resistances shunting the junctions of the random Josephson junction array. The first film is $\rm 40 \AA$, and the thickness increment is $\rm 10 \AA$.

\noindent{\bf Figure 5} Fits of the conductance of the insulating films (taken at 10 K) and the normal state conductance of the superconducting films (also taken at 10 K) to 2D percolation function, for Bi films on Ge.  Only the critical thicknes in each regime is different, as discussed in the text. 

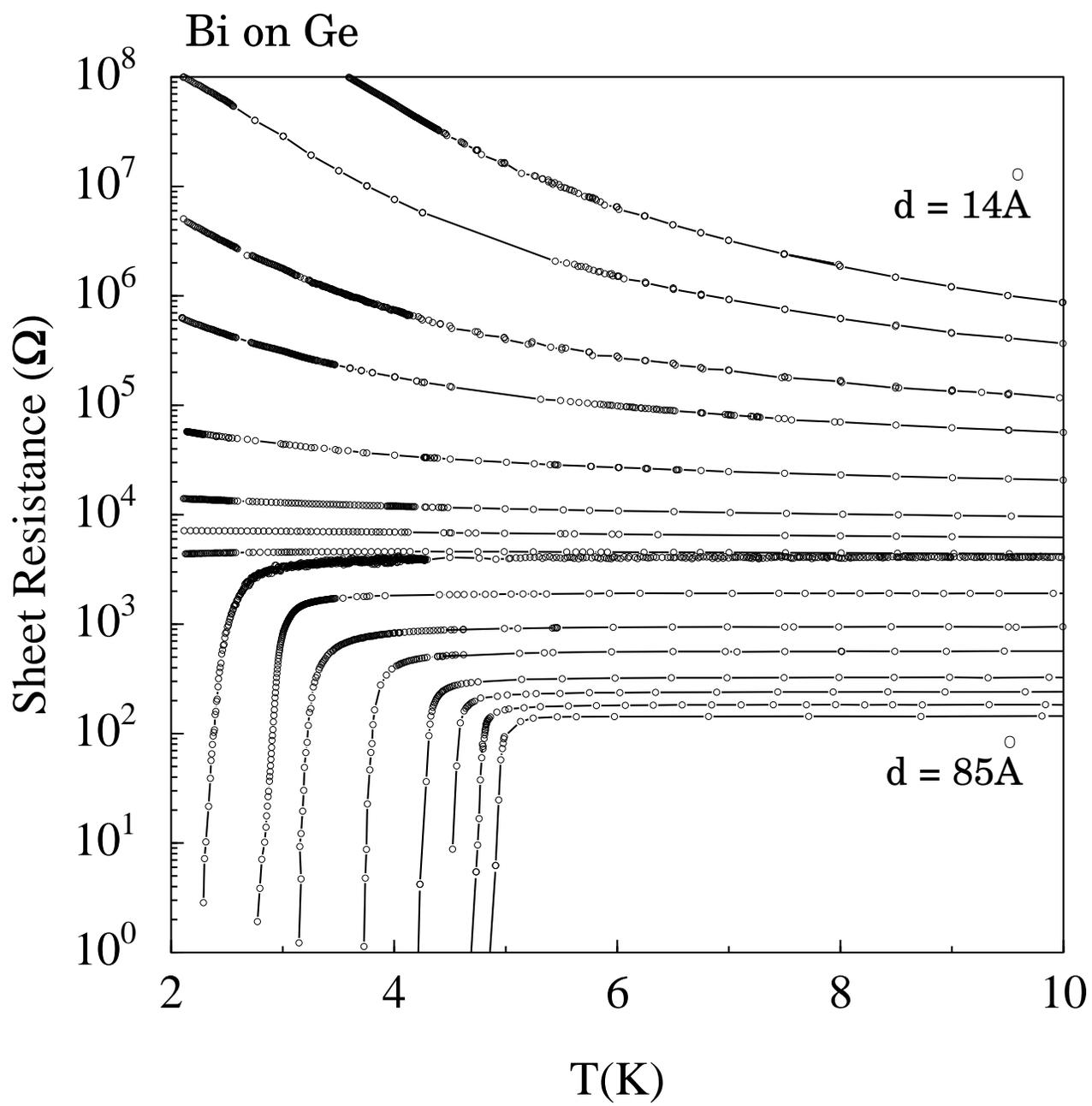
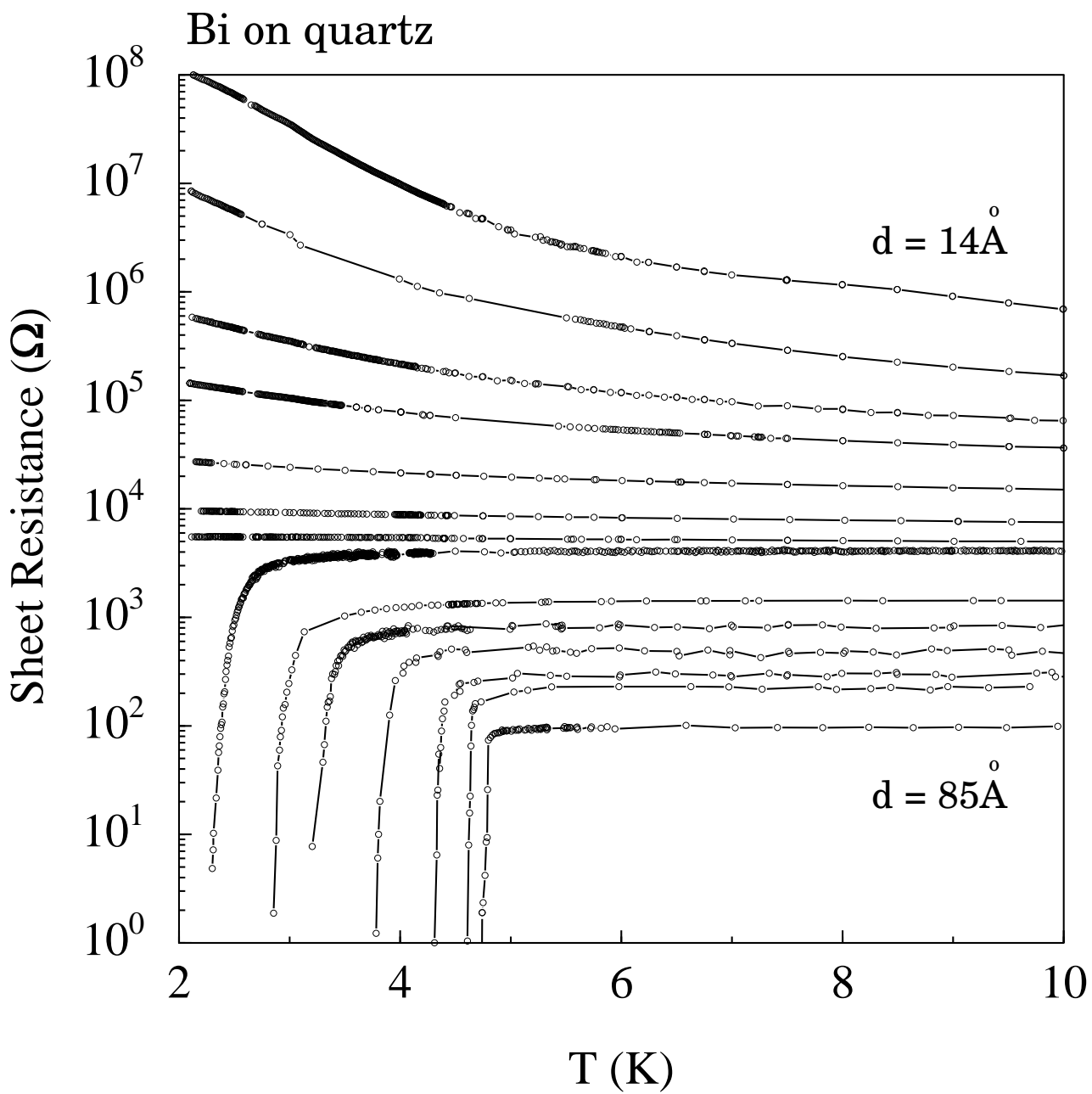
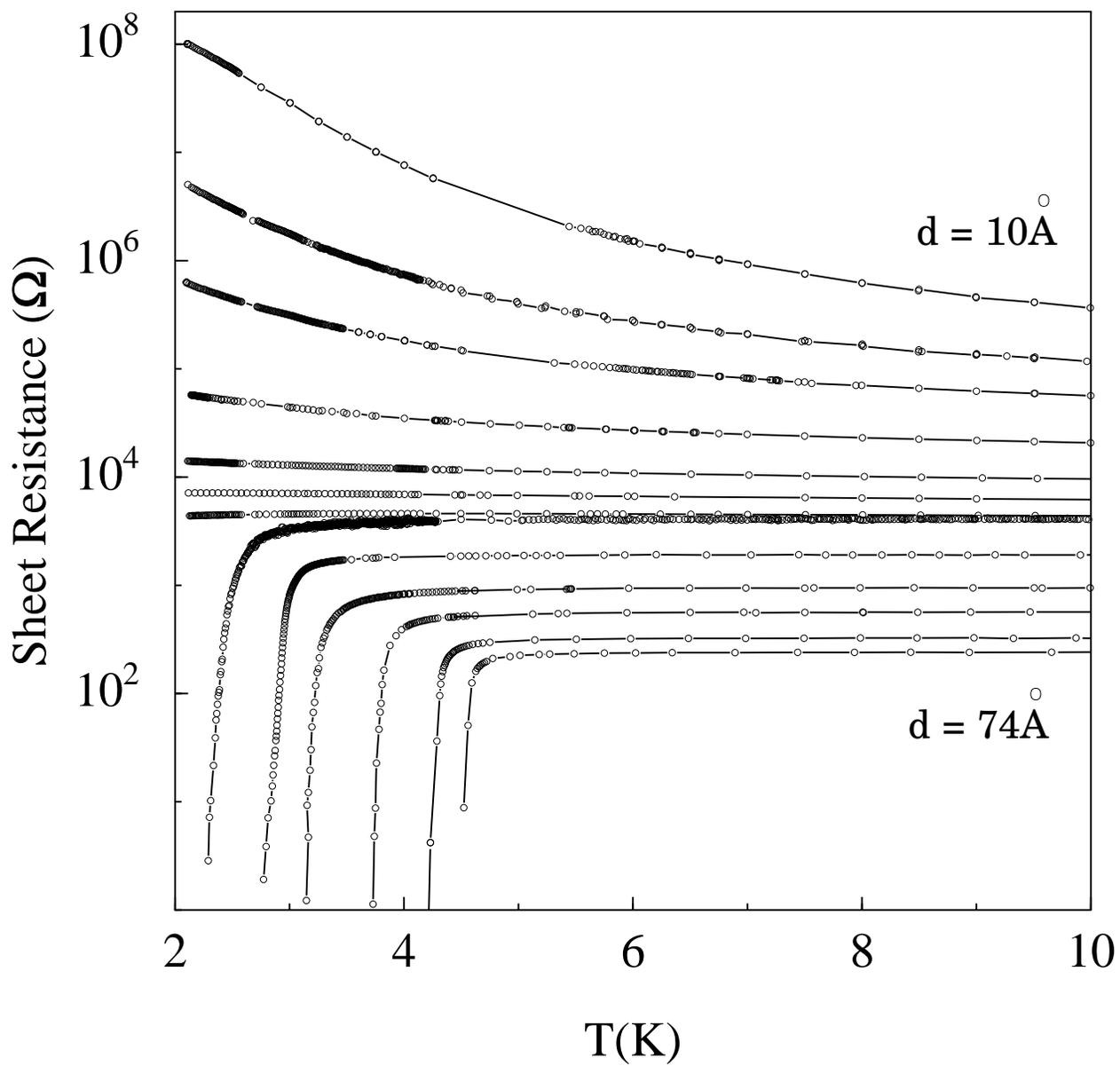
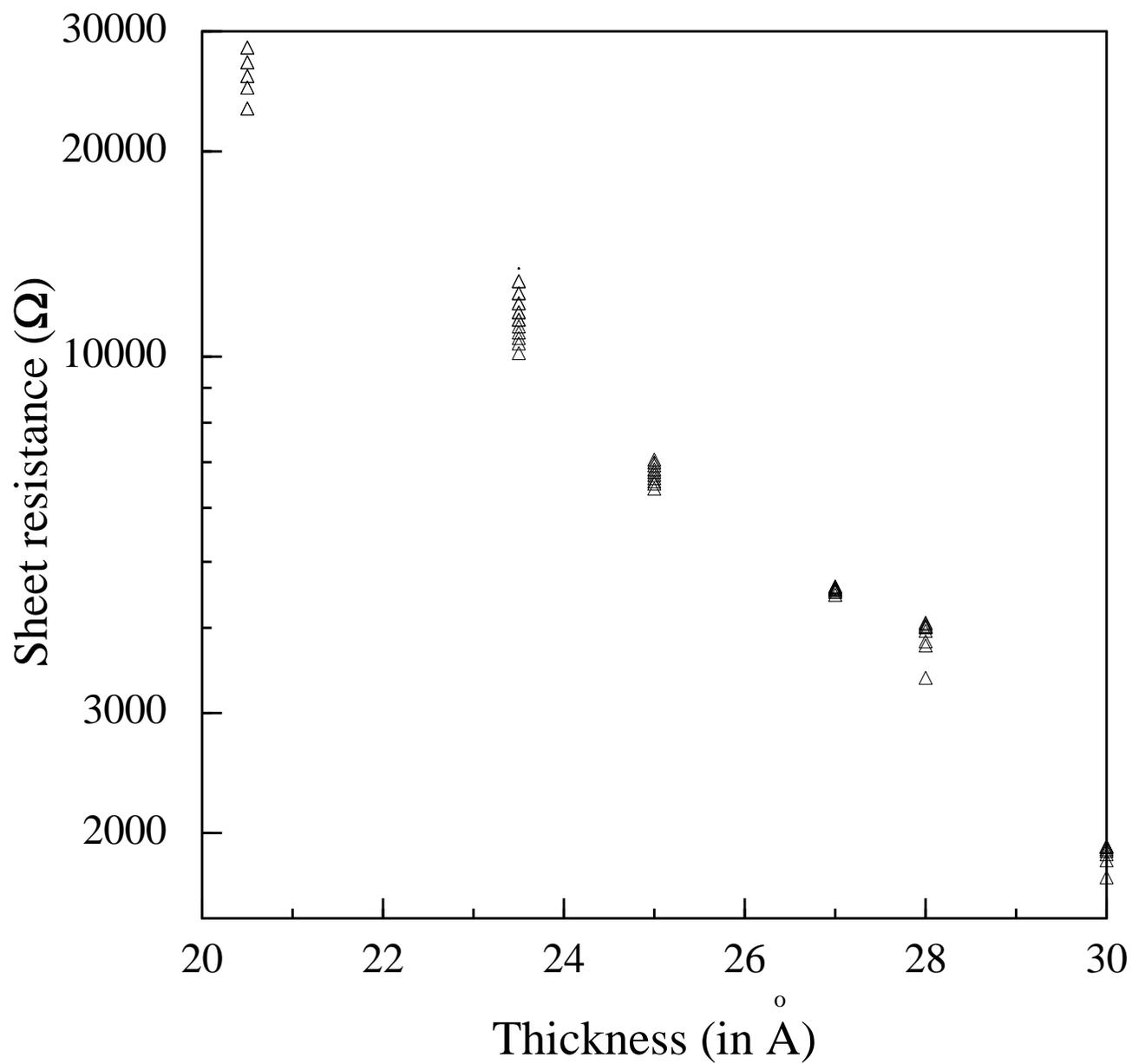
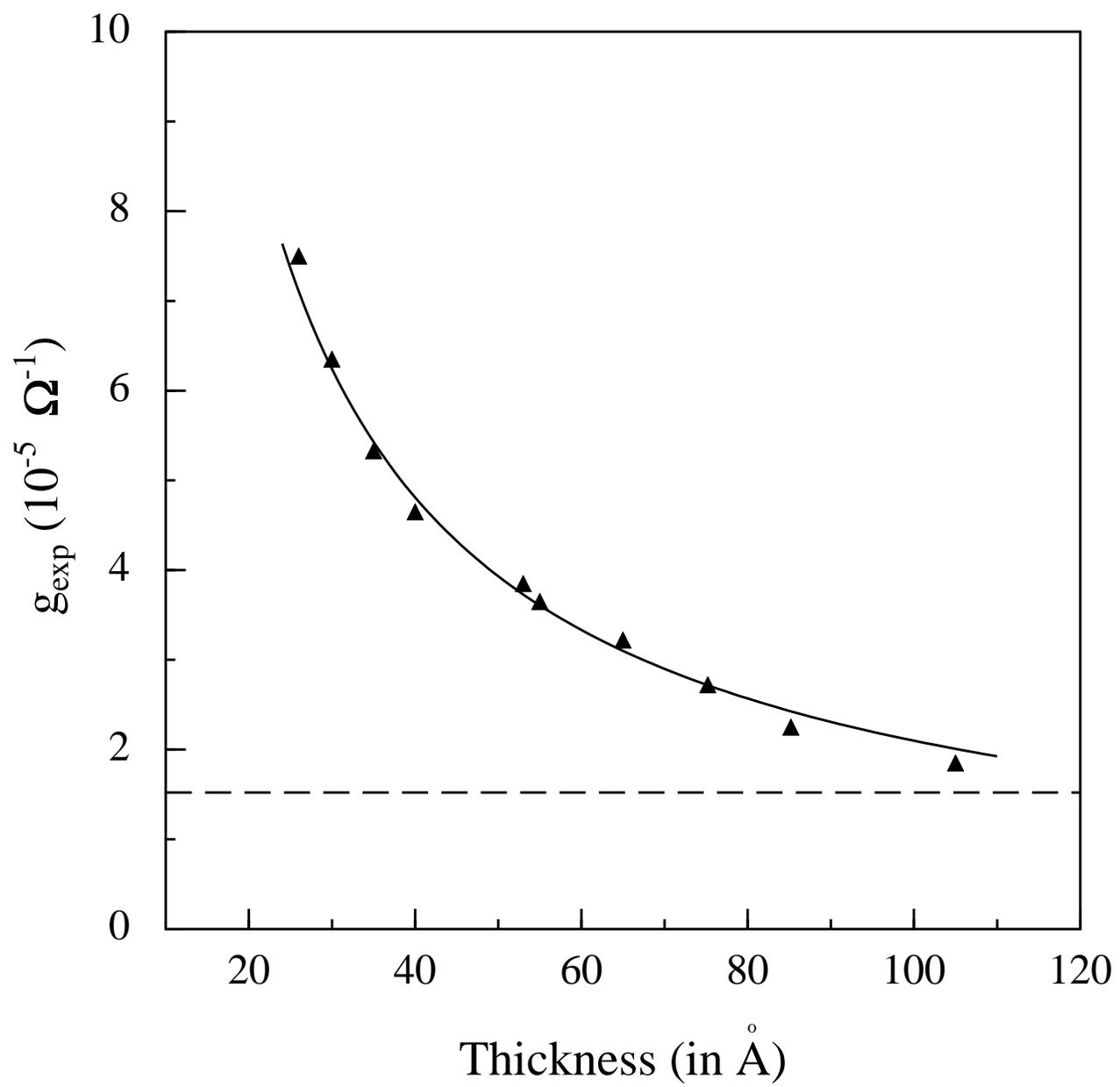
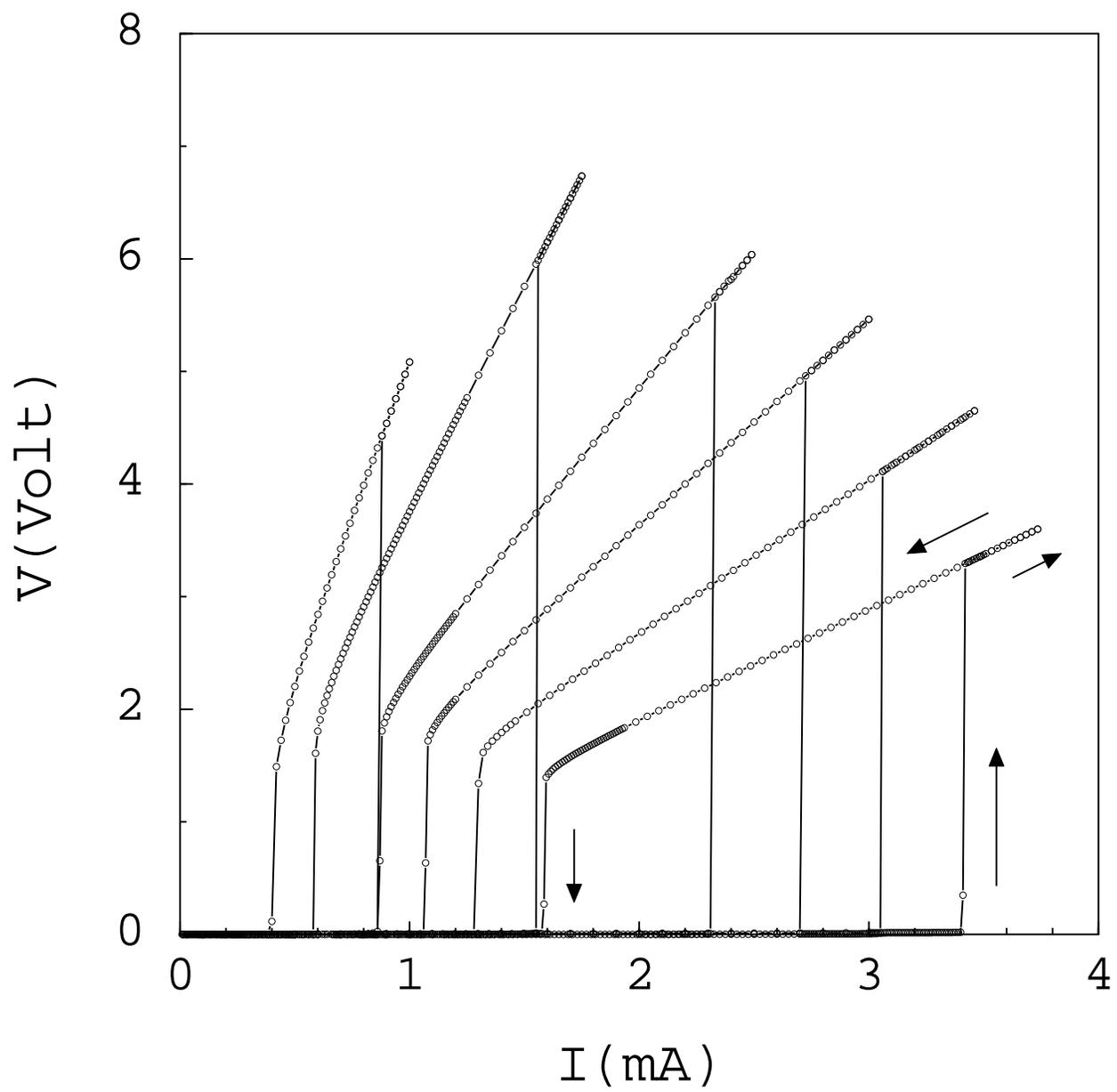
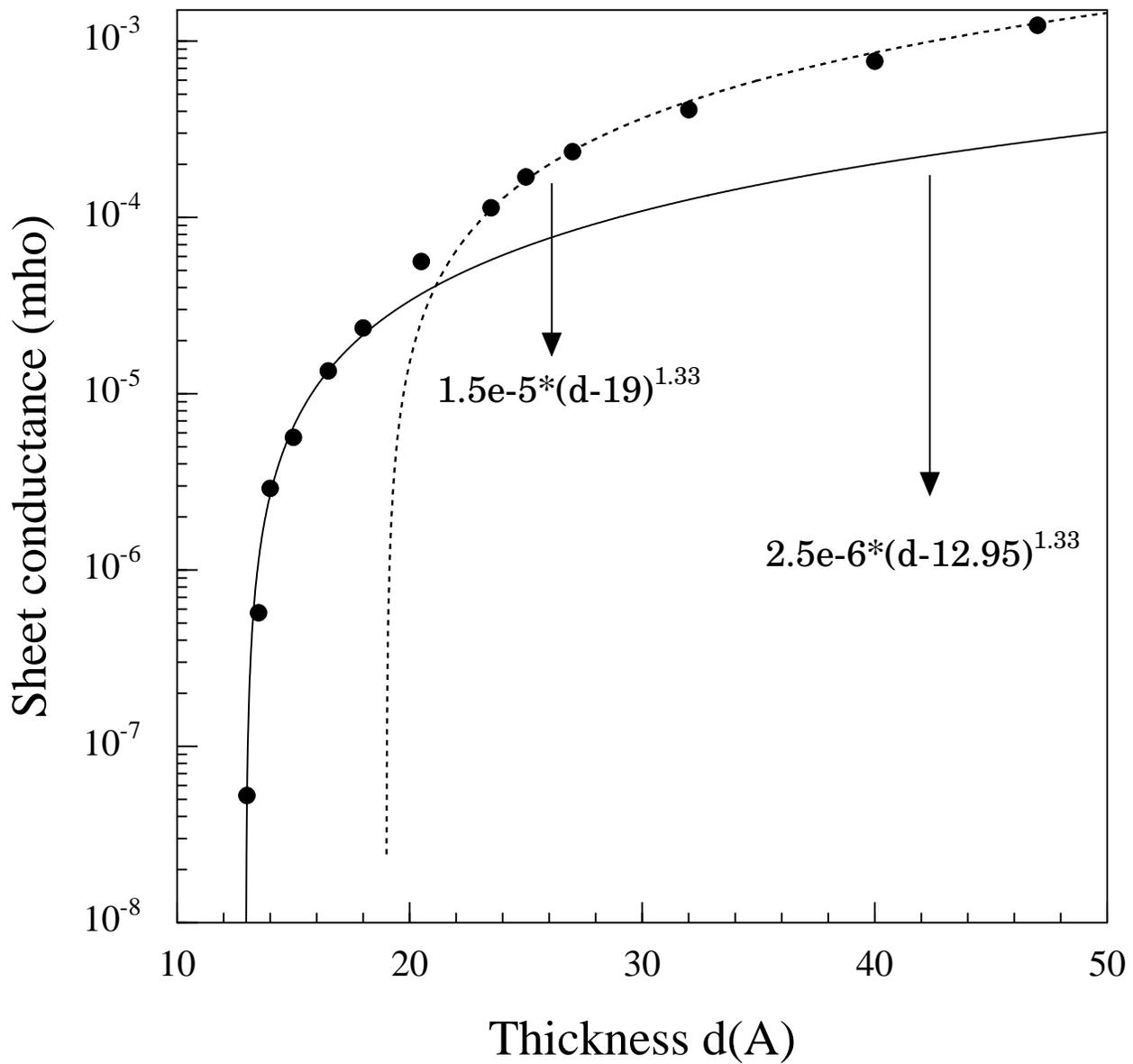
\end{document}